\documentstyle[12pt,aaspp]{article}
\begin{document}

\title{MAGNITUDE OFFSET BETWEEN LENSED STARS AND OBSERVED STARS --
A NEW PROBE OF THE STRUCTURE OF THE GALACTIC BAR}

\author{K.~Z.~Stanek\altaffilmark{1}}
\affil{Princeton University Observatory, Princeton, NJ 08544--1001}
\affil{\tt e-mail I: stanek@astro.princeton.edu}
\altaffiltext{1}{On leave from N.~Copernicus Astronomical Center,
Bartycka 18, Warszawa 00--716, Poland}

\begin{abstract}

We propose a new method that can be used to constrain the properties of
the Galactic bar (bulge). If the majority of the lensing objects are in the
Galactic bar, then we predict a systematic offset in the apparent magnitude
between lensed stars and all observed stars. Using OGLE color-magnitude
diagram data we model this effect in the region of the diagram dominated by
bulge red clump stars and find that for some models of the Galactic bar the
expected offset in the apparent magnitude may be as large as $0.2\;mag$.
About 100 lensed stars in the red clump region of the color-magnitude diagram
is needed to unambiguously detect this effect, a number within the reach of
current microlensing projects. We find a good correlation between the extent
of the bar along the line of sight and the expected magnitude offset. We also
obtain a constraint for the extent of the bar along the line of sight using the
observed luminosity function for the red clump stars.

\end{abstract}


\section{INTRODUCTION}

The OGLE project (Optical Gravitational Lensing Experiment)
recently announced the results from the first two years
of the search for gravitational microlensing in the direction of the
Galactic bulge (Udalski et al.~1994). A total of nine lenses were found
in the combined data from 1992 and 1993, giving an estimate of the
optical depth to gravitational microlensing of the Galactic bulge stars:
$\tau= 3.3 \pm 1.2 \times 10^{-6}$ for Baade's Window and the nearby
Galactic Bar fields. Results of the MACHO experiment (Alcock et al.~1994)
give a similar, high value for the optical depth in the bulge. Theoretical
models predicted a lower value, somewhere in the range of $ 0.5 - 1.0
\times 10^{-6}$ (Paczy\'nski 1991; Griest et al.~1991; Kiraga and
Paczy\'nski 1994; Giudice et al.~1994). Alcock et al.~(1994), and also
Gould~(1994), argue that the extra mass responsible for the observed high event
rate may be located in the Galactic disk. However, in considering the results
from Udalski et al.~(1994), Paczy\'nski et al.~(1994b) show that the lenses
unlikely occur in the Galactic disk, due to the current constraints on the
mass in the Galactic disk (Kuijken \& Gilmore 1991; Bahcall, Flynn \& Gould
1992) and the observed ``hole'' in the disk (deficiency of stars as compared to
exponential disk, Paczy\'nski et al.~1994a). Instead, they argue for the
lenses to be located in the Galactic bar, which is oriented towards us.
Calculations by Zhao, Spergel \& Rich (1994b) of the microlensing by the bar
stars are consistent with the arguments in Paczy\'nski et al.~(1994b).
Here, we propose and discuss a simple observational method that allows us to
distinguish between the two possibilities. The test we propose is a local
one, i.e. it does not require observations in fields with different Galactic
longitudes (Kiraga \& Paczy\'nski 1994; Paczy\'nski et al.~1994b; Evans 1994;
Kiraga 1994).

There is now a number of photometric and dynamical indications that the
Galaxy is barred (de Vaucouleurs 1964; Blitz \& Spergel 1991;
Binney et al.~1991; Whitelock \& Catchpole 1992; Weinberg 1992;
Blitz 1993). The bar shows in the OGLE color-magnitude diagram data as a many
sigma difference of $\sim 0.37\;mag$ in the apparent magnitude of the
red clump stars in the two opposite $l=\pm5\deg$ OGLE Galactic bar fields
(Stanek et al.~1994). It is also clearly seen in COBE data (Weiland et
al.~1994), which was used by Dwek et al.~(1994) to constrain a number of the
bar models existing in the literature. In this paper we will use results of
Dwek et al.~(1994), combined with {\tt ftp}-accessible color-magnitude OGLE
data (Udalski et al.~1993; for {\tt ftp} instructions see Paczy\'nski et
al.~1994a) to investigate a new effect of the gravitational microlensing.
If the majority of the lensing objects are in the Galactic bar, then there
should  be a systematic offset in the apparent magnitude between observed
stars and lensed stars -- stars from the far side of the bar are much more
likely to be lensed than the stars from the near side of the bar. We discuss
various aspects of the magnitude offset between the observed stars and lensed
stars and show that it can be used as a new probe of the structure of the
Galactic bar (bulge).

\section{THE DATA}

We use the color-magnitude diagram (CMD) of Udalski et al.~(1993a)
for one of the nine fields in Baade's Window (BW3, coordinates of the center
of the field: $l=0.92\deg, b=-4.19\deg$). This is so far the only field
observed by OGLE in which stars from the red clump region were lensed.
The observations were made using the 1 meter Swope telescope at the
Las  Campanas Observatory, operated by the Carnegie Institution of Washington,
and $2048\times 2048$ Ford/Loral CCD detector with the pixel size 0.44 arcsec
covering $15' \times 15'$ field of view. The CMD for BW3 field can be seen in
Udalski et al.~(1993). Most of the diagram is dominated by bulge stars, with
a distinct red clump, red giant, and turn-off point stars. The part of the
diagram dominated by disk stars for this and other BW fields was analyzed by
Paczy\'nski et al.~(1994a). Stanek et al.~(1994) used well-defined
population of bulge red clump stars in nine BW fields and four Galactic Bar
fields to find an evidence for the Galactic bar. Red clump stars are
the equivalent of the horizontal branch stars for a metal rich population,
i.e. relatively low mass stars burning helium in their cores. From observations
and also from stellar evolution theory (Castellani, Chieffi \& Straniero 1992)
we expect the bulge red clump stars to be relatively bright and have a narrow
luminosity distribution, with weak dependence on the  metallicity. Therefore,
red clump stars form a suitable population with which to investigate the
properties of high-metallicity systems, like the  Galactic bulge.

To analyze the distribution of bulge red clump stars in a quantitative
manner, we use the extinction-insensitive $V_{_{V-I}}$ parameter
(Paczy\'nski et al.~1994a; Stanek et al.~1994)
\begin{equation}
  V_{_{V-I}} \equiv V - 2.6 ~ (V-I),
\label{eq:free}
\end{equation}
where we use reddening law $ E_{_{V-I}} = A_{_V}/2.6 $, following Dean,
Warren, \& Cousins (1978) and Walker (1985). The parameter $V_{_{V-I}}$
has been defined so that if $A_{_V}/E_{_{V-I}}$ is independent of location
then for any particular star its value is not affected by the unknown
extinction (see Stanek et al.~1994). Then we consider only the
region of the CMD for BW3 field clearly dominated by the bulge red clump stars:
\begin{equation}
1.4  < V-I ~~;~~ 10.5 < V_{_{V-I}} < 14.0
\label{eq:select}
\end{equation}
Stars observed in the BW3 field that satisfy the inequalities
(\ref{eq:select}) were counted in bins of $\Delta V_{_{V-I}} = 0.05 $. The
result appears in Fig.1, where we see the number of stars as a function of
$V_{_{V-I}}$. Also shown are the values of the $V_{_{V-I}}$ parameter for the
two lenses observed in this field near the red clump
($V_{_{V-I}}=12.51,13.11$).

The region selected above corresponds roughly to stars with $I<18$. For such
stars it was found by Udalski et al.~(1993, their Table 4), from
artificial star test, that the completeness of the sample is rather uniform,
on the level of 0.8, which makes our further analysis much simpler. Also, red
clump stars provide us with very convenient ``landmark'', a point which will
be used in the next section.

\section{THE MODEL}

We want to reproduce the observed luminosity function (Fig.1) assuming
one of Dwek et al.~(1994) best-fit  models of the Galactic bar density
distribution and fitting some simple form of the intrinsic luminosity
function for the stars in the red clump region.

For the Galactic bar density distribution we take Dwek's et al. G2 model with
the exponential cutoff at 2.4 kpc. This model is described by
analytical formula
\begin{equation}
\rho_{G2}(x,y,z)=\rho_0 \exp(-r_s^2/2)\;\left[M_{\odot}\,pc^{-3}\right],
\end{equation}
where $r_s$ is defined as
\begin{equation}
r_s\equiv\left\{\left[\left(\frac{x}{x_0}\right)^2+\left(\frac{y}{y_0}\right)^2
\right]^2+\left(\frac{z}{z_0}\right)^4\right\}^{1/4}.
\end{equation}
The scale lengths of the bar are $x_0=1.47,~y_0=0.56,~z_0=0.42~\;kpc$
(assuming the distance to the center of  the Galaxy $8.0\;kpc$)
and the angle of inclination to the line of sight is $20\deg$
(in Dwek's et al. nomenclature this corresponds to $\alpha=70\deg$).
The parameter $\rho_0$ is given by a total mass of the bar
by $\rho_0=M_{B}/(8\pi x_0\, y_0\, z_0)$.
For the $M_{B}\approx 2\times10^{10}\;M_{\odot}~\rho_0$ is equal to $\sim2.3\;
M_{\odot}\;pc^{-3}$.
We parametrize the fitted luminosity function in the form
\begin{equation}
\Phi(L)=\left(\frac{N_0}{L_{\odot}}\right)\left(\frac{L}{L_{\odot}}
\right)^{-\alpha}+\frac{N_{RC}}{\sigma_{RC}\sqrt{2\pi}}\exp\left[-
\frac{(L-L_{RC})^2}{2\sigma_{RC}^2}\right]~\;
\left[L^{-1}_{\odot}\right].
\end{equation}
The power-law with index $\alpha$ is intended to represent the underlying broad
population of stars, and the Gaussian is intended to represent red clump stars.
For a given $V_{_{V-I}}$ bin, the number of stars in that bin is then given by
\begin{equation}
N(V_{_{V-I}})=C_1\int_{D_{MIN}}^{D_{MAX}}\rho(D_s)D_s^2\Phi(L)L\;dD_s,
\label{eq:integ}
\end{equation}
where we took $\rho(D_s)=\rho_{G2}(x,y,z)$,
$D_{MIN}=3\;kpc,~D_{MAX}=13\;kpc,~C_1$ is a constant, $D_s$ is a
distance from the observer to the source, and
$L=C_2 D_s^2\; 10^{-0.4 V_{_{V-I}}}$,
where $C_2$ is another constant. The choice of $D_{MIN}$ and $D_{MAX}$
is not important as long as they contain the distance cutoff applied
to the density distribution. In equation~(\ref{eq:integ}) we assume
that the number of observable stars is everywhere proportional to the
density of matter and that the intrinsic luminosity function is independent
of location, which gives a constant $(M/L)$ throughout the bar.

We then fit the observed luminosity function using the above
assumptions. The fit is shown in Fig.1.
The fit is satisfactory, considering the fact that we use an assumed model
of the density distribution, which is not the subject to the fit.
The peak of the fitted intrinsic luminosity function for
the red clump stars is very narrow, with fitted $\sigma_{RC}$ corresponding to
$\sim 0.07\;mag$ (although luminosity function $\Phi(V_{_{V-I}})$ is no longer
Gaussian), a point which we will address in the discussion.

Having the intrinsic luminosity function, we can now obtain
the predicted luminosity function for lensed stars, in the case where
the lenses are located in the bar with the described density distribution.
In this case the equivalent to equation~(\ref{eq:integ}) will have the form
\begin{equation}
N_L(V_{_{V-I}})=C_1\int_{D_{MIN}}^{D_{MAX}}\tau(D_s)\rho(D_s)
D_s^2\Phi(L)L\;dD_s,
\label{eq:integlen}
\end{equation}
where $\tau(D_s)$ is the local optical depth for gravitational microlensing
given by the formula
\begin{equation}
\tau(D_s)=\frac{4\pi G}{c^2}\int_0^{D_s}\rho(D_d)\frac{D_d(D_s-D_d)}{D_s}
\;dD_d,
\end{equation}
$D_d$ being the distance from the observer to the deflector.
The resulting luminosity function for the lensed stars, multiplied
by the average optical depth for microlensing for all stars,
is shown in Fig.1. Clearly, the expected luminosity
function for lensed stars is different from the luminosity function for
all observed stars. This is because the stars on the far side of the bar
are more likely to be lensed than the stars on the near side of the bar,
a purely geometrical effect.

Knowing that we should expect differences in the observed luminosity
functions of all stars and of lensed stars, we want to ask the question:
how many lensed stars in the red clump region of the CMD do we need
to tell with a desired accuracy that the two functions are really
different? We checked a number of simple tests and various statistics
and found that the median of the sample is a good measure of the difference
in the case when we restrict the region of comparison to $\pm 1 \;mag$ from
the peak of the observed luminosity function. More quantitatively --- we take
a running region $2\; mag$ wide. Then for 100 lenses drawn randomly 10,000
times from both the fitted  luminosity function and from the predicted
luminosity function of lensed stars, we compare the parameter
$\kappa=(\left<med_2\right>-\left<med_1\right>)/
(\sigma_{med,2}+\sigma_{med,1})$,
i.e. we find a region in $V_{_{V-I}}$, of a given width, where we are
maximally sensitive to the difference between the two distributions.
Subscript $1$ corresponds to all stars, subscript $2$ to lensed stars.
The value of $\kappa$ as a function of the middle point of the selected
region is shown in the panel A of Fig.2. With the thick horizontal line
we mark the level of $\kappa=1.5$. We see that $\kappa$ has relatively
flat-topped distribution, so as to include as many predicted lenses into
region as possible without loosing much sensitivity, we choose the region
between $11.2 < V_{_{V-I}} <13.2$. Then, for this region we draw a random
sample of $n$ lenses many times from each distribution and we compare
the corresponding $\left<med\right>$ and $\sigma_{med}$. The plot
of these values for each distribution as a function of $\log(n)$ is shown
in the panel B of Fig.2. The mean difference between the medians of the
two distributions is $\Delta mag=0.185\;mag$. Also shown
is the median value of the $V_{_{V-I}}=12.81$ parameter for the two lenses
observed by OGLE so far in the red clump region. This value suggests, although
with low significance, that the lensing objects may be in the Galactic bar.
Visual analysis of the published color-magnitudes diagrams with the positions
of lensed stars by MACHO experiment (Alcock et al.~1994) shows that two out
of four events for one of the fields have positions close to, but below the
red clump.

\section{DISCUSSION}

In the previous section we have shown that in the case when the
majority of microlensing objects are the Galactic bar, there should be
an offset in the median magnitude between all observed stars (in the bar)
and the lensed stars. We now discuss the implications of this conclusion.

The size of the magnitude offset between the observed and the
lensed stars depends on how elongated the bar (bulge) is along
the line of sight. We define an elongation parameter for the bar to be
\begin{equation}
\left(\frac{\Delta l}{2}\right)^2\equiv\frac{\int_{D_{MIN}}^{D_{MAX}}
(D_s-\left<D_s \right>)^2\rho(D_s)\;dD_s}{\int_{D_{MIN}}^{D_{MAX}}
\rho(D_s)\;dD_s}.
\label{eq:elong}
\end{equation}
We take three models of the bar (bulge) from the existing literature, one
axially-symmetric model from Kent~(1992), and two models from Dwek et
al.~(1994), noted by G2 (used in this paper) and E3. All these models were
subject to the same distance cutoff, discussed earlier in this paper.
First we construct the intrinsic luminosity function by fitting the data
for the BW3 field. Fitted luminosity functions have $\sigma_{RC}$ that
correspond to $\sim 0.07\;mag$  for G2 model, $\sim0.35\;mag$ for E3 model, and
$\sim0.25\;mag$ for Kent's model. We then calculate, for each of these
models, both $\Delta l$ and the expected magnitude offset $\Delta mag$ for
the line of sight towards the BW3 field. The results are
shown in Fig.3. There is a good correlation between $\Delta l$ and
$\Delta mag$, which indicates that the method discussed in this paper may
also be used to constrain the elongation of the bar along the line of sight,
a constraint that is otherwise difficult to obtain. To some degree such a
constraint is already put by inspecting the observed luminosity
function of the red clump stars (Fig.1). When using G2 model
with $2.4\;kpc$ density cutoff and angle of inclination of $20\deg$, we were
able to fit the observations only with very narrow ($\sigma_{RC}\sim0.07\;mag$)
intrinsic luminosity function for the red clump stars.
When trying to fit the observed luminosity function using Dwek's et al. G2
model with $5\;kpc$ cutoff and angle of inclination to the line of sight of
$13\deg$ (one of Dwek's et al. best-fit models), the overall fit was worse,
and even with the red clump stars distribution approaching perfect standard
candle ($\sigma_{RC}=0$), the resulting fitted luminosity function was too
broad. This model has the elongation along the line of sight, as defined by
Eq.\ref{eq:elong}, $\Delta l=2.65\;kpc$, which we may therefore treat as
approximate upper limit for the elongation of the bar along the line of sight
towards BW3. As it should be possible to obtain independently the width
of the intrinsic luminosity function for the red clump stars, either from
the stellar evolution theory or from the observations of high-metallicity
clusters, this, already interesting, upper limit of $\Delta l<2.65\;kpc$ can
be eventually translated to the purely photometric estimate of the $\Delta l$
value.

We also checked how the efficiency of the proposed method depends on the
Galactic coordinates of the observed field. For number of fields with
different galactic coordinates we investigated the value of $\kappa$,
analogously to what was done for the BW3 field. We found no significant
differences in the maximum value of $\kappa$
one can obtain, the only difference was the expected shift in the peak
$V_{_{V-I}}$ value of the red clumps stars distribution due to the bar
geometry.

The lensing events of the stars on the far side of the bar are not only more
probable, but should also be longer lasting, for geometrical reasons. However,
the timescale of the event is not only a function of how big the Einstein
ring of the lensing object is, but it also depends on the velocities of
both the lens and the source, which is difficult to quantify without a
kinematical model of the stars in the bar (see Zhao~1994, used by Zhao et
al.~1994b). One may however expect that the average timescale of the observed
events $\left< t_0 \right>$ will be a function of $V_{_{V-I}}$ in the red clump
region of the CMD, and in the case when geometry is most important for the
timescales, lensed stars that are fainter should have on average longer
lasting microlensing events.

The lensed stars in the vicinity of the red clump region provide us
with an important opportunity in studying the kinematics  of the stars
in the Galactic bar. As Zhao, Spergel \& Rich (1994a) mention, it would be
very useful to obtain a sample of stars with the distances known with accuracy
sufficient to place the star on either the near or far side of the bulge.
But as we have shown, the lensed stars in the red clump region of the CMD
may be very likely on the far side of the bulge, and they are bright
enough to be found on old photographic plates to determine their proper
motions.

In order to unambiguously detect the effect discussed in this paper,
we estimate the number of required lensed stars in the red clump CMD region
be about $\sim 100$. For the OGLE project it translates to a total of
$\sim 500$ observed microlensing events. For the MACHO group, which because
of poorer seeing has a brighter limiting magnitude, the total number may be
factor of $\sim 2$ smaller. These numbers are certainly within the reach of
current or next generation microlensing experiments.

\acknowledgments{I would like to thank Bohdan Paczy\'nski for suggesting
the topic of this paper and for many stimulating discussions. I also thank
Neil Tyson, Wes Colley and David Spergel for useful comments on earlier
version of this paper. Part of this work was done when at the Royal Greenwich
Observatory, which I thank for the summer studentship. This project was
supported with the NSF grant AST 9216494 and the NSF Grant-in-Aid of Research
through Sigma Xi, The Scientific Research Society.}

\newpage

\begin{figure}

\begin{center}
{\bf FIGURE CAPTIONS}
\end{center}

\caption{Observed number of stars as a function of the $V_{_{V-I}}$ parameter
for BW3 OGLE field ($l=0.92\deg, b=-4.19\deg$) is shown, with error-bars
corresponding to square-root of $N$ in each bin. For the purpose of
presentation, the data were box-car smoothed with the bin width equal to 3
and only every third point is shown. Vertical long-dashed lines correspond
to the values of the $V_{_{V-I}}$ parameter for the two lenses observed in this
field near the red clump. Also shown, with the  continuous thick line, is
the fit to the observed luminosity function, and with the dotted thick line,
the expected luminosity function for lensed stars.}

\caption{Panel A shows the efficiency $\kappa$ of the magnitude offset
measurement as a function of mid-range point for $2\;mag$ wide running region
(see text). Panel B shows the average magnitude and standard deviation
for the randomly drawn lenses from the fitted luminosity function
(lower points) and expected luminosity function for lensed stars (upper
points) as a function of number of the lenses in the sample $n_{lens}$.
The black dot corresponds to the median of $V_{_{V-I}}$ for the two
lenses observed so far by OGLE in the red clump region.}

\caption{Correlation between the elongation of the bar (bulge) model
along the line of sight $\Delta l$ and the expected magnitude offset
$\Delta mag$ between all observed stars and lensed stars. The dashed
line is shown only to guide the eye.}

\end{figure}


\begin{references}

\reference Alcock, Ch.~et al.~1994, ApJ, in press (SISSA preprint
astro-ph/9407009)

\reference Bahcall, J.~N., Flynn, C., \& Gould, A.~1992, ApJ, 389, 234

\reference Binney, J., Gerhard, O.~E., Stark, A.~A., Bally, J., \&
Uchida, K.~I., 1991, MNRAS, 252, 210

\reference Blitz, L.~1993, in ``Back to the Galaxy'', AIP Conference
Proceedings 278, eds.~S.~S.~Holt \& F.~Verter, (New York: AIP Conference
Proceedings), p.98

\reference Blitz, L.,  \& Spergel, D.~N., 1991, ApJ, 379, 631

\reference Castellani, V., Chieffi, A.~\& Straniero, O., 1992, ApJS, 78, 517

\reference Dean, J.~F., Warren, P.~R., \& Cousins, A.~W.~J., 1978,
MNRAS, 183, 569

\reference Dwek, E., et al. 1994, ApJ, submitted

\reference de Vaucouleurs, G.~1964, in IAU Symp.~No.~20, eds.~F.~J.~Kerr
and A.~W.~Rodgers (Australian Acad.~Sci.: Canberra), p.195

\reference Evans, N.~W.~1994, ApJ, in press (SISSA preprint astro-ph/9409002)

\reference Giudice, G.~F., Mollerach, S., and Roulet, E.~1994, Phys.~Rev.~D.,
in press (SISSA preprint astro-ph/9312047)

\reference Gould, A., 1994, ApJ, submitted (SISSA preprint astro-ph/9408060)

\reference Griest, K.~et al.~1991, ApJ, 372, L79

\reference Holtzman, J. A., et al., 1993, AJ, 106, 1826

\reference Kent, S.~M.~1992, ApJ, 387, 181

\reference Kiraga, M., 1994, Acta Astron., submitted

\reference Kiraga, M., \& Paczy\'nski, B.~1994, ApJ, 430, L101

\reference Kuijken, K., \& Gilmore, G.~1991, ApJ, 357, L9

\reference Paczy\'nski, B.~1991, ApJ, 371, L63

\reference Paczy\'nski, B., Stanek, K.~Z., Udalski, A., Szyma\'nski, M.,
Ka\l u\.zny, J., Kubiak, M., \& Mateo, M.~1994a, AJ, 107, 2060

\reference Paczy\'nski, B., Stanek, K.~Z., Udalski, A., Szyma\'nski, M.,
Ka\l u\.zny, J., Kubiak, M., Mateo, M., \& Krzemi\'nski, W., 1994b,
ApJ, 435, L113

\reference Stanek, K.~Z., Mateo, M., Udalski, A., Szyma\'nski, M.,
Ka\l u\.zny, J., Kubiak, M., 1994, ApJ, 429, L73

\reference Udalski, A., Szyma\'nski, M., Ka\l u\.zny, J., Kubiak, M.,
\& Mateo, M.~1993, Acta Astron., 43, 69

\reference Udalski, A., Szyma\'nski, M., Stanek, K.~Z.,
Ka\l u\.zny, J., Kubiak, M., Mateo, M., Krzemi\'nski, W.,
Paczy\'nski, B., \& Venkat, R.~1994b, Acta Astron., 44, 165

\reference Walker, A.~R., 1985, MNRAS, 213, 889

\reference Weiland, J.~L., et al., 1994, ApJ, 425, L81

\reference Weinberg, M.~D., 1992, ApJ, 384, 81

\reference Whitelock, P., \& Catchpole, R., 1992, in The Center, Bulge,
and Disk of the Milky Way, ed.~L.~Blitz, (Dordrecht: Kluwer Academic), p.~103

\reference Zhao, H. S., 1994, Ph.~D.~thesis, Columbia University

\reference Zhao, H.~S., Spergel, D.~N., \& Rich, M.~R.~1994a, AJ, in press
(SISSA preprint astro-ph/9409024)

\reference Zhao, H.~S., Spergel, D.~N., \& Rich, M.~R.~1994b, ApJ, submitted
(SISSA preprint astro-ph/9409022)

\end{references}
\end{document}